# Community-Based AI Learning: Redistributing Artificial Intelligence's Epistemic Authority in Education

Santiago Ojeda-Ramirez
School of Education
University of California, Irvine
Irvine, California, USA
sojedara@uci.edu

Symone Gyles
School of Education
University of California, Irvine
Irvine, California, USA
sgyles@uci.edu

Kylie Peppler
School of Education
University of California, Irvine
Irvine, California, USA
kpeppler@uci.edu

## ABSTRACT

As generative AI systems increasingly mediate learning, they are often treated as authoritative sources of knowledge. This perspective paper introduces community-based AI learning as a framework that repositions authority, grounding AI engagement in learners' lived and community-based epistemologies. Drawing from community-driven learning and constructionist traditions, we articulate three commitments: epistemic fine tuning, redistribution of authority, and situated discernment. Together, these processes localize critical AI literacy by calibrating trust, foregrounding community knowledge, and supporting collective judgment about when to design with, interrogate, or reject AI. We argue that equitable AI education requires negotiating authority through place, history, and social context.

• Social and professional topics → Computing literacy

• Human centered computing → Participatory design

## 1 Introduction

Artificial intelligence and machine learning (hereafter, AI/ML) are rapidly expanding beyond research laboratories and becoming embedded in everyday life, increasingly shaping how knowledge is produced, accessed, and interpreted. Students and teachers encounter these systems not only as tools they actively use, but also as infrastructures that influence what information is surfaced, how explanations are generated, and which forms of knowledge are rendered visible or authoritative. In response, scholars and policymakers have called for the widespread development of AI literacy, emphasizing the need for learners to understand, use, and critique AI applications [24, 47]. Yet, much of the work on AI in education has centered on the design of AI as instructional agents—such as artificial teachers, tutors, or co-learners—implicitly positioning AI systems as credible sources of explanation and guidance, and learners as recipients or collaborators within AI-mediated knowledge practices [34]. This framing has become increasingly consequential with the intensification of generative AI technologies in education settings: young people now interact with large language models on a daily basis, curriculum providers are rapidly reshaping educational offerings, and educators face mounting pressure to integrate AI/ML content into classrooms. Amid this acceleration, comparatively little attention has been paid to how AI systems come to be treated as legitimate knowers, and how students' community-based, lived, and localized ways of knowing are positioned in relation to these systems.

As generative AI systems become embedded in everyday learning practices, their outputs are frequently delivered with confidence, fluency, and an authoritative tone. These interface features, combined with widespread institutional adoption, can position AI systems as legitimate, and at times definitive, sources of knowledge within classrooms [5], providing them with epistemic authority. *Epistemic authority* refers to who or what is treated as a credible knower: whose claims are trusted, deferred to, and used to determine what counts as valid knowledge. Importantly, epistemic authority is not neutral [9], it is historically structured through relations of power. Scholars of epistemic injustice have shown how credibility is unevenly distributed, often marginalizing community-based, Indigenous, and Global South epistemologies in favor of dominant Eurocentric knowledge formations [26, 42]. Recent research in AI ethics and social epistemology demonstrates that generative AI systems reproduce these asymmetries: Large language models are trained on globally scraped datasets that disproportionately reflect Western, English-dominant, and Global North perspectives [10]. Empirical studies document how these systems encode and amplify racialized, gendered, and cultural biases [4]. As a result, when students engage AI as a knowledge source, they are engaging epistemologies that are already unevenly structured. Equity-oriented AI education has begun to interrogate this dynamic. Work on youth auditing of AI systems [28], critical interrogation of AI outputs, and computational empowerment [16] positions learners as epistemic agents who exercise judgment, authorship, and decision-making power in relation to computing systems. While these approaches productively redistribute technical agency, they often stop short of centering learners' communities as primary epistemic resources in interpreting AI systems.

In this perspective paper, we argue that there is a need to further interrogate the contextuality of AI as an epistemic authority by explicitly connecting students' lived experiences in their communities. As such, we present community-based AI learning as a framework for pairing AI learning with and around community interest and use. Drawing from community-based science education [12], we define community-based AI learning as an approach to AI education in which learners engage with AI systems through issues, practices, and forms of knowledge grounded in their communities, treating local experience and lived realities as central resources for learning with and about AI. Furthermore, we draw on constructionist learning to theorize how interrogations of AI authority can be externalized through design and making, positioning the construction of AI-mediated artifacts as a central site for negotiating knowledge, authority, and relevance. From this perspective, community-based AI learning offers a needed framework for redistributing epistemic authority, and positioning learners and their communities as legitimate knowers alongside and in dialogue with AI systems. This paper advances this perspective by articulating the theoretical foundations of community-based AI learning, situating constructionism as a lens for learning with and about AI, contrasting this approach with dominant AI-as-authority framings in computing education, and outlining its implications for equity-oriented and critically situated AI literacy.

## 2 Background

### 2.1 Dominant Paradigms in AI Education

Research on computing education and the learning sciences has increasingly explored the use of AI systems as instructional supports, including AI tutors, conversational agents, and AI-based scaffolds designed to guide learners through problem-solving, explanation, and feedback processes [34]. The purpose of learning with and about AI is largely framed in instrumental terms: to support learners in mastering disciplinary content, completing tasks more effectively, or collaborating productively with AI systems. In education, the central aim is often improvement: better feedback, more personalized guidance, optimized learning trajectories.

Although such designs provide meaningful support, they tend to position AI outputs as provisional but legitimate knowledge contributions. By supplying explanations, examples, and evaluative feedback, AI tools often function as default reference points for what counts as correct, relevant, or complete knowledge—particularly in contexts where learners lack prior expertise. As Cooper [5] cautions, generative AI systems risk positioning themselves as "ultimate epistemic authorities," (p. 449) presenting responses that appear comprehensive and confident even when they are partial, decontextualized, or incorrect. Recent empirical work further shows that students frequently treat ChatGPT-generated content as epistemically trustworthy, integrating both correct and incorrect information into academic work, especially when their epistemic beliefs privilege authority-based knowledge sources [48]. Even after verification strategies, the broader orientation assumes AI as a resource to be consulted, calibrated, or corrected.

### 2.2 Epistemic Authority in AI-Mediated Learning

Epistemic authority refers to who or what is treated as a legitimate source of knowledge: whose claims are trusted, deferred to, and used to decide what counts as knowing. Nevertheless, authority is unevenly distributed, structured by social power, race, class, and global positionality. These same sociocultural forces shape AI systems, which increasingly mediate instructional processes—not only as tools, but as contributors to explanation, evaluation, and sense-making. In educational settings, this participation subtly reorganizes where epistemic weight is placed. Dominant approaches to AI education position AI systems as central reference points that organize judgments about correctness, relevance, and completeness. The question is, therefore, not simply whether AI outputs are accurate, but how authority is configured when algorithmic systems enter pedagogical spaces.

Work in social epistemology reminds us that authority is always socially negotiated. Who is granted credibility, whose interpretations are prioritized, and whose knowledge is treated as foundational are outcomes shaped by historical and political arrangements. As Fricker [9] argues, individuals can be wronged in their capacity as knowers when their credibility is unjustly discounted due to prejudice. Decolonial scholarship further demonstrates that modern knowledge infrastructures have long elevated certain epistemologies—particularly Eurocentric and Global North traditions—while subordinating local, Indigenous, and community-rooted ways of knowing [26, 36] . These hierarchies are not abstract; they are materially embedded in institutions and systems.

Contemporary AI research shows that such hierarchies are reproduced through training data, model architectures, and evaluation benchmarks. Generative systems are built upon large-scale corpora that disproportionately reflect dominant linguistic and cultural formations [10]. Empirical studies show that these models learn, reproduce, and amplify social biases along axes of race, gender, and culture, thereby reinforcing existing epistemic hierarchies, not abstracting from them [52]. Related work has further argued that generative AI can exacerbate epistemic injustice by privileging dominant cultural frames while obscuring or distorting the interpretive resources of marginalized communities [4]. Thus, epistemologies embedded in generative AI systems align with globalized visions of knowledge that present themselves as general or universal, while systematically overlooking the situated, community-based, and lived dimensions of knowledge. When learners encounter AI-generated explanations, they are engaging outputs shaped by these structural asymmetries.

What distinguishes this form of epistemic authority from others that equity-oriented education has historically contested is its scale and its performed universality. AI speaks to millions of learners with the same voice regardless of their community or history, presenting responses as what Haraway [13] calls a "view from nowhere." Crawford [8] shows that this universality is a fiction: AI encodes specific data, labor, and power arrangements that consistently reflect dominant cultural formations. Community-based AI learning is the tailor-made contestation to this specific form of authority, in the same way that community-driven science positioned itself as a direct response to extractive dominant science [3].

### 2.3  Equity-Oriented Reorientations

In response, equity-oriented scholarship has expanded the purpose of AI learning beyond performance optimization. Computational empowerment frames computing education as cultivating critical judgment and participatory agency in relation to technology [16]. Critical AI literacy emphasizes examining the social and political implications of AI systems [50]. For example, some work has engaged youth in auditing AI-powered technologies, surfacing how algorithmic systems reproduce social biases and inequities while positioning students as critical agents capable of interrogating AI systems [28]. Other scholarship frames AI outputs as tentative contributions that require verification and contextual judgment, noting that while chatbots can scaffold technical problem solving, they do not replace learners' responsibility for domain-specific reasoning and ethical discernment [41].

Similar sociopolitical commitments appear in RESPECT conference research that foregrounds inequity in generative AI, such as moving 'beyond the hype' of AI toward ethical and societal implications [33], and articulates justice-oriented AI education through concepts such as data sovereignty [29]. These approaches bring necessary attention to power, bias, and justice in AI learning. Yet, the sociopolitical is often treated at a broad or generalized scale, with less sustained attention to how learners' local, lived, and community-based knowledge can function as a primary epistemic resource for interpreting and re-authoring AI systems. Importantly, research on youth-as-AI-designers highlights how engaging learners in designing, training, and testing AI systems can foster agency, ethical awareness, and deeper understanding [18]. When students are positioned as designers, design necessarily involves decisions about what problems matter, whose values are encoded, and what knowledge is taken up by AI systems. Drawing from design justice [7], such decisions are inseparable from centering the knowledge and experience of those most directly impacted by the design. Community members bring expert knowledge grounded in lived experience that governs what problems get named, how AI is evaluated, and what counts as a good outcome.

Thus, a question that remains insufficiently theorized is not only how learners use, analyze or design AI systems, but *how their communities function as primary sites of knowledge in determining AI's meaning and legitimacy*. Recent work by Akgün and colleagues [2] grounds critical AI literacy in students' cultural assets, funds of knowledge, and critical consciousness about AI's societal implications, through participatory design with youth and community partners in informal K-8 settings. Community-based AI learning builds on this foundation while making a different epistemological claim: that community knowledge holds interpretive authority over AI outputs, functioning as the evaluative standard rather than a cultural resource for enrichment. It is this gap that community-based AI learning seeks to address.

## 3  Theoretical Foundations

### 3.1  Community-Based Learning as a Situated and Cultural Practice

Community-based learning emerges from critiques of educational models that treat knowledge as fixed, universal, and transferable independent of learners' social worlds. Across domains, traditional instruction has often emphasized "ready-to-use" knowledge—canonical explanations, standardized representations, and decontextualized practices—positioning learning as the reproduction of disciplinary content rather than as participation in meaningful activity [51]. In contrast, sociocultural and social constructivist perspectives conceptualize learning as a situated, relational process in which meaning is constructed through interaction, dialogue, and engagement in culturally and historically shaped practices [38]. From this view, learning cannot be understood only as a cognitive acquisition but necessarily as a social process that unfolds through participation in communities and shared activity [11, 21].

Community-driven approaches further extend this perspective by foregrounding the political and ethical dimensions of learning. Ballard and colleagues argue that community-driven science and education exist at the "edges" of equity, justice, and disciplinary learning, where learners navigate tensions between normative disciplinary knowledge and the lived realities of their communities. From this stance, communities are not sites where disciplinary knowledge is merely applied, but places where knowledge is produced, contested, and reinterpreted in relation to local concerns and collective histories [3]. Learning, therefore, is oriented not only toward understanding disciplinary concepts, but toward developing the capacity to use knowledge to make sense of and intervene in issues that matter in learners' lives.

Within this broader orientation, Gyles & Clark articulate community-based science learning (CBS) as a framework for instruction anchored in locally and socially relevant phenomena, where *community* extends beyond geographic boundaries to include cultural epistemologies, historical ontologies, and social structures shaping lived experience [12]. CBS explicitly rejects the notion of science—or learning more broadly—as neutral or context-free. Instead, it conceptualizes learning as a collection of social experiences shaped by power, history, and identity. A central contribution of CBS is its redefinition of expertise: rather than positioning teachers or disciplinary texts as the sole epistemic authorities, CBS foregrounds heterogeneous expertise, recognizing students' community knowledge, family histories, and everyday experiences as legitimate and necessary resources for making sense of phenomena. Learning environments informed by CBS are thus designed to redistribute epistemic authority, creating space for negotiation, co-construction, and shared meaning-making.

Two pedagogical traditions also center community in learning, yet make a different kind of claim. Culturally relevant pedagogy [20] positions community cultural knowledge as a resource for academic achievement and critical consciousness. Culturally sustaining pedagogy [37] goes further, arguing that community practices deserve preservation as intrinsic expressions of democratic pluralism, as ends worth sustaining in their own right, independent of their utility for reaching dominant cultural destinations. Community-based learning asks a different question: who has the authority to say what counts as true, relevant, and worth knowing about the conditions that shape community life. The answer is that community members hold robust forms of situated local knowledge about those conditions [17], and that learning must be organized around that knowledge as its foundation, treating community knowing as the standard against which other sources, including AI, are evaluated [3, 12].

This orientation resonates strongly with parallel work in computer science education that centers community, belonging, and equity. Ryoo and colleagues demonstrate how community-based approaches in computing—such as family and community engagement, professional learning communities, and youth-driven collectives—can dismantle deficit narratives and reposition learners as capable knowledge producers within computing practices [44]. Across these studies, computing learning is conceptualized as socially situated and collaboratively constructed, where inquiry unfolds through relationships among teachers, youth, families, and local communities, and where identities, histories, and collective concerns shape participation in meaningful ways [44, 45]. These perspectives offer a robust theoretical foundation for extending community-based approaches to emerging domains such as AI learning, where questions of epistemic authority, legitimacy, and whose knowledge counts are increasingly consequential.

Community-based science learning and community-centered computing education show that disciplinary rigor, equity, and community expertise can be co-constitutive within instructional design. This redistribution of authority, in which community-rooted ways of knowing organize inquiry, grounds our articulation of community-based AI learning.

### 3.2 Learning (with Technology) as Constructing Models and Artifacts to Share in Community

Constructionism has been a foundational approach for understanding both the affordances and the limits of learning *with* and *about* technology. As a learning theory and pedagogical orientation, constructionism posits that learning happens most effectively when learners construct models both *in their mind*s and *in the world*, and when they externalize their emerging understanding through the creation of shareable artifacts [19, 35]. These artifacts can take many forms—"a poem, a sandcastle, or a computer program" [36] —and function as public entities through which learners test ideas, make meaning visible, and invite reflection and dialogue. Constructionism thus emphasizes learning as a process of *building knowledge through making*, where thinking is materially and socially mediated.

This perspective has been widely used to understand learning with different technologies in at least two ways. First, technologies can function as media for constructing public artifacts meaningful to learners and their communities. Work on culturally responsive making with American Indian girls demonstrates how e-textiles can braid Indigenous sewing traditions with circuitry and programming, repositioning cultural knowledge as central to computational learning [46]. Research on Scratch in informal urban settings has similarly shown how youth use visual programming to design videogames and interactive art that remix pop culture and neighborhood spaces, positioning programming as a medium for reauthoring dominant media genres through locally meaningful cultural references [38]. In the context of AI, recent work has extended this by positioning generative AI as a medium for speculative design, enabling youth to craft prompts and visual artifacts that materialize community-rooted futures and externalize collective concerns about how AI might reshape their local worlds [32].

Second, constructionism has also been mobilized to understand learning through the construction and deconstruction *of technologies themselves*. Work on youth coding and app design demonstrates that when young people learn to code and build digital tools with an explicit civic or cultural purpose, they develop agency and authorship while cultivating critical awareness of how technologies shape representation, participation, and power [22]. In AI education, this orientation is evident in work that engages high school students in building small generative language models, or "babyGPTs," by intentionally curating datasets, training models, and evaluating outputs [27]. Across these approaches, construction means that students are producing functional artifacts while being engaged in reflective and critical relationships with technology as something that can be shaped, questioned, and reimagined.

If community-based learning is conceptualized as learning grounded in issues from learners' own communities, constructionism allows us to argue that stronger mental models of what AI is, how it can be used and applied, and how it can be evaluated are developed when students create something in the world that is relevant to their communities and grounded in their lived experiences. This involves both using AI as a medium to create a public entity for their communities, and creating, conceptualizing, and thinking about AI systems that are themselves grounded in community issues. Through these processes, learners externalize their understanding of AI in ways that are socially situated, open to critique, and responsive to local concerns— activating the specific affordances that community-based learning environments make possible. In this way, constructionism offers not only a complementary learning theory but also a mechanism for enacting community-based AI learning by making community knowledge an active resource in shaping and contextualizing AI systems.

## 4 Community-Based AI Learning as a perspective

Building from community-based science and community-driven learning [3, 12], we define community-based AI learning as an approach to learning with and about AI that anchors epistemic activity in locally meaningful concerns, redistributes authority across heterogeneous forms of expertise, and orients AI engagement toward collective sense-making and action in relation to lived conditions.

Grounding learning in community knowledge in this way rests on two axiological justifications. The first is epistemic justice [9]: when AI holds epistemic authority in a learning space, community ways of knowing are discounted in favor of outputs that do not reflect students' lives or places. The second is that community knowledge is the kind of knowing AI cannot produce: knowing built through living in a place, through intergenerational relationships, through embodied community life [8, 13]. As Paris [37] grounds culturally sustaining pedagogy in the democratic value of sustaining cultural pluralism, community-based AI learning is grounded in the value of sustaining situated, embodied, and community-rooted knowing as the foundation from which human judgment about AI becomes possible [17].

This perspective does not treat AI knowledge as abstract or universally transferable. Instead, it situates AI within the cultural, political, historical, and infrastructural realities that shape how communities encounter, use and understand data, platforms, and algorithmic systems. Community-based AI learning rests on three interrelated commitments:

First, AI learning requires situated epistemic negotiation. Learners do not simply acquire accurate mental models of AI systems. Instead, they develop evolving judgments about what AI is, when it is trustworthy, and where its limits lie in connection with their own lived experience. AI outputs, therefore, are engaged as materials for interpretation and deliberation, not as definitive conclusions as students recalibrate trust, delimit use, and bring community knowledge into conversation with algorithmic outputs. In this sense, learning involves what we later call epistemic fine-tuning: not adjusting AI/ML models, but adjusting one's stance toward them. This commitment extends constructionist principles by grounding model building in phenomena that matter locally, and it extends community-based learning by positioning AI knowledge as something negotiated between technical representations and lived expertise.

Second, AI learning must redistribute epistemic authority. Community-based AI learning refuses the configuration of AI as ultimate authority and, instead, positions learners as legitimate knowers of their communities and the conditions under which AI operates. Expertise is heterogeneous: it includes technical understanding, experiential knowledge, cultural interpretation, and civic judgment. Thus, authority becomes relational and conditional as learners evaluate when AI assistance is appropriate and when it should be resisted, supplemented, or rejected. This redistribution does not dismiss AI. Instead, it reframes it as one voice among many within a community of inquiry. In doing so, it localizes critical AI literacy by asking whose knowledge counts, who benefits, and who bears harm in specific places and contexts in relation to AI.

In practice, this redistribution reorients AI education away from student-machine epistemic relationships and toward human-community epistemic relationships. AI outputs are evaluated through dialogue with people who hold situated local knowledge grounded in specific places and conditions [17]. This reorientation cultivates what Pleasants and colleagues [39] call technoskepticism: a disposition toward continuously examining the relationships communities want to have with technology, asking whose needs AI serves, and whether its presence produces more wellbeing or more harm.

Third, AI engagement centers collective discernment and action. Learning with AI is not confined to skill acquisition or abstract critique. It becomes a process through which communities examine how AI infrastructures intersect with their social worlds and decide how to respond. In some contexts, this may involve designing with AI to address local concerns. In others, it may require building against AI systems that exacerbate injustice. Criticality, therefore, is not uniform across settings. It is shaped by geography, infrastructure, exposure to harm, and local histories of technological entanglement. Community-based AI learning makes space for differential criticality by place, acknowledging that the ethical and political stance toward AI cannot be predetermined in advance.

These commitments position community-based AI learning as a framework that integrates constructionist making, community-based epistemologies, and critical AI literacy into a situated theory of learning. Within this framework learners refine their epistemic relationships to AI, redistribute authority between systems and communities, and deliberate collectively about whether, how, and why AI should participate in their shared futures.

In practice, this looks like a unit in which students design something real for their community: a restaurant concept, a neighborhood tool, a local cultural archive. They use AI as a design resource while treating community knowledge as the evaluative standard. A visit to a community site precedes any AI engagement. Community practitioners serve as co-evaluative judges of student work. The making process externalizes not just students' design ideas but their negotiated relationship to AI itself.

## 5 Community-Based AI Learning in practice

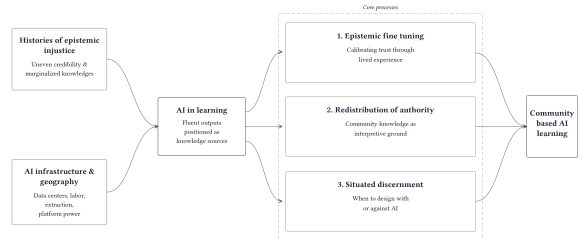

Figure 1: Community-Based AI Learning: From AI-Centered Authority to Collective Discernment.

### 5.1 Epistemic Fine Tuning

In technical ML usage, fine tuning refers to updating a pretrained model's parameters with additional task or domain data so the model performs better for a specific purpose. Contemporary Natural Language Processing and Large Language Models work has popularized this framing through transfer learning pipelines in which a base model is adapted for downstream tasks by fine tuning [14], building on earlier neural network transfer research that already described a pretraining stage followed by "fine tuning" weights for a new task [40]. In education adjacent AI work, fine tuning increasingly appears as a design strategy for making models better at teaching through conversations [43, 49]. These strands share an assumption: fine tuning improves the model.

Epistemic fine tuning is not a model adaptation, but related in spirit. We use the term to describe a learner-centered calibration of how AI outputs are interpreted, trusted, and used when learning is anchored in community concerns. Through this process, the learner's epistemic relationship to the model is recalibrated: learners treat AI claims as partial and decontextualized, then actively adjust their confidence by comparing outputs against lived experience, local histories, community expertise, and the consequences of acting on that information. This involves drawing on community grounded knowledge as an interpretive resource, clarifying when AI systems are appropriate and when they are not, and treating trust of AI outputs as conditional and context dependent. Gyles and Clark [12] show that students positioned as knowledge constructors in community-based science develop the capacity to evaluate disciplinary claims through their own lived experience. In AI learning, this means a student who knows their neighborhood's economic conditions, food culture, or family history is equipped to notice when an AI output describes a place that does not match the place they actually live in, and to say so, and to explain why.

Decolonial scholarship helps clarify why this epistemic move matters: coloniality has historically positioned Eurocentric knowledge as universal while relegating localized ways of knowing to the margins, producing systematic hierarchies of credibility and legitimacy [42]. Epistemic fine tuning operationalizes a reversal of that hierarchy in everyday AI mediated learning by treating community knowledge as a primary authority for sense making, not a secondary "example" after AI speaks.

This move is complementary to approaches that engage youth as designers and auditors of AI, including work where learners build simplified models and critically examine outputs and data practices [28]. Epistemic fine tuning can be supported by such designs, but it is not restricted to them. It can also occur whenever learners are invited to interrogate AI responses through community grounded criteria of relevance, harm, and usefulness, and to continually revise what they take AI to be "good for" in the first place.

### 5.2 Redistributing Authority and Localizing Critical AI Literacy

Community-based AI learning seeks to precisely counter this dynamic by situating authority within communities whose lived experiences, histories, and local conditions exceed and at times contest model outputs, positioning AI systems as one interpretive input among others. Jadallah and Ballard [17] found that community members brought knowledge of specific watersheds, land histories, and local ecological conditions that professional scientists did not hold. When community-based AI learning is organized around similar relationships, when a family member's account of the neighborhood, a practitioner's design judgment, or a community elder's historical knowledge shapes what students accept or question in an AI output, the learning space itself is reorganized around who actually knows the conditions at hand.

Critical AI literacy, from this perspective, cannot remain abstract or universally defined. For example, Alemán's [1] counter-cartographies of AI demonstrate how communities experience AI infrastructures differently, particularly when data extraction, predictive policing, or automated decision systems materially shape everyday life. AI is not encountered uniformly across geographies; it is embedded in local infrastructures, labor arrangements, and surveillance regimes. Cortéz et al.'s [6] AlgoRitmo literacies similarly illustrates how algorithmic systems are interpreted and contested through culturally specific practices and artistic interventions. These works suggest that criticality is not a generic stance toward bias or fairness; it emerges through situated engagements with how AI operates in a particular community. Community-based AI learning therefore treats sociopolitical analysis as anchored in material and cultural context, not as a generalized checklist of harms.

Work on youth as designers of AI further supports this orientation. Iivari et al.[15] show that young people can design AI systems for socially meaningful purposes, engaging ethical and civic considerations in context-sensitive ways. Community-based AI learning extends this insight by emphasizing that such design must be accountable to local epistemologies and consequences [7]. At the same time, scholarship advocating a return to Luddism [23,30] and literacies in platform ecologies [31] reminds us that not all critical engagement entails building with AI. In some contexts—particularly those disproportionately harmed by data extraction or infrastructural inequities—criticality may require refusal, redesign, or resistance. Decisions to engage, redesign, or decline AI systems are grounded in local knowledge, shared experience, and collective judgment, with technological progress understood as something to be evaluated in context.

## 6 Implications

For computing educators, this perspective calls for a recalibration of how AI systems operate within classroom knowledge practices. Educators can create structured opportunities for students to evaluate AI outputs using locally grounded criteria of relevance, harm, and usefulness, positioning AI as one interpretive resource within collective sensemaking. This approach does not require abandoning technical rigor or dismantling standards-aligned curricula. It introduces practices of epistemic calibration in which learners compare AI-generated claims with community knowledge, lived experience, and shared values. Although community-based pedagogies have been critiqued for demanding substantial curricular redesign and teacher reorientation [25], community-based AI learning can function modularly. It can coexist with technical instruction while adding a layer of ethical and interpretive discernment, particularly when AI intersects with socially meaningful domains.

The concrete forms this takes are epistemologically consequential. When students engage community members who hold situated local knowledge [17], visiting sites where their communities gather, consulting practitioners whose expertise is grounded in lived local experience, presenting designs to people who can evaluate them against local conditions, they place community knowledge at the center of interpretation. These human-to-human relationships cultivate what Pleasants and colleagues call technoskepticism [39]: a disposition to continuously examine what relationships communities want with technology, asking whose needs AI serves and whether its presence produces more wellbeing or harm. That evaluation may lead to conditional use, shaped by community criteria, or to refusal, when community knowledge reveals that AI misrepresents or harms the people it claims to serve.

For curriculum designers and policy makers, this perspective shifts critical AI literacy [50] from generalized awareness of sociopolitical issues of AI toward situated engagement with AI consequences for a specific community. Curricula, for example, can move beyond addressing abstract bias examples to examine AI as it appears in students' social worlds, including surveillance systems, recommendation platforms, predictive tools, and cultural production technologies [2]. Policy initiatives that frame AI competencies as universally necessary for workforce or civic participation [24] can also benefit from incorporating contextual interpretation of AI's role within local educational ecologies. Importantly, in communities disproportionately affected by data extraction, environmental burdens, or algorithmic harm, critical engagement may involve limitation, refusal, or strategic non-use; in other settings, AI may support culturally grounded design and collective problem-solving; guided everytime.by ethical standards of use.  For researchers, this orientation positions *situatedness*: place, infrastructure, and lived experience as constitutive dimensions of AI learning. Interestingly, community-based AI learning can generalize across contexts: through comparison of specific community issues but specially through the practice of epistemic discernment: learners of many communities becoming capable of recalibrating trust in AI systems, negotiate their authority, and evaluate them in relation to local conditions. Methodologically, this invites community partnerships, participatory designs, and analytic approaches that treat geography context as central to understanding how AI learning unfolds [2].

## 7 Conclusion

This paper argues that community-based AI learning reframes what it means to learn with and about artificial intelligence. Community-based AI learning names a specific epistemic harm: an AI that speaks as if from everywhere and nowhere, encoding dominant cultural formations while presenting itself as universal [8, 13]. Its antidote is grounding AI engagement in the robust situated local knowledge of human communities [17], organized around human-community relationships and oriented toward the question of what relationships communities want to have with technology [39]. As AI systems become infrastructural and unevenly distributed, equity cannot rest on generalized competencies or abstract critique alone. If AI increasingly mediates knowledge claims, the central question becomes: *who calibrates that authority and under what conditions?* Community-based AI learning anchors that responsibility within the communities who experience its consequences.

## 8 Positionality statement

These commitments arise from our intersecting trajectories as community-engaged scholars. Santiago, a Latino scholar and former K–12 teacher working with Latinx youth, centers how young people draw on cultural knowledge and lived experience when engaging with AI. Symone, a Black woman community-based science education researcher and former middle school teacher, situates STEM and AI learning within their social, historical, and political contexts, advancing communities as knowledge authorities. Kylie, a learning sciences scholar grounded in creative production and long-term partnerships with minoritized communities, approaches AI as a sociotechnical system whose authority can be contested through locally grounded practice.

## 9 Declaration of AI use

During the preparation of this work the authors used Claude in order to proofread the text. After using this tool/service, the authors reviewed and edited the content as needed and take full responsibility for the content of the published article.